\documentclass[sigconf]{acmart}

\def\BibTeX{{\rm B\kern-.05em{\sc i\kern-.025em b}\kern-.08emT\kern-.1667em\lower.7ex\hbox{E}\kern-.125emX}}
    
\copyrightyear{2018}
\acmYear{2018}
\setcopyright{acmlicensed}
\acmConference[Woodstock '18]{Woodstock '18: ACM Symposium on Neural Gaze Detection}{June 03--05, 2018}{Woodstock, NY}
\acmBooktitle{Woodstock '18: ACM Symposium on Neural Gaze Detection, June 03--05, 2018, Woodstock, NY}
\acmPrice{15.00}
\acmDOI{10.1145/1122445.1122456}
\acmISBN{978-1-4503-9999-9/18/06}

\acmSubmissionID{sp379}

\usepackage{enumerate}
\usepackage{upgreek}
\usepackage{tabularx}

\begin{document}

%
\title{A study on the Interpretability of Neural Retrieval Models using DeepSHAP}

%
\author{Zeon Trevor Fernando}
\affiliation{%
  \institution{L3S Research Center}
  \streetaddress{}
  \city{Hannover}
  \country{Germany}}
\email{fernando@l3s.de}

\author{Jaspreet Singh}
\affiliation{
  \institution{L3S Research Center}
  \city{Hannover}
  \country{Germany}}
\email{singh@l3s.de}

\author{Avishek Anand}
\affiliation{
  \institution{L3S Research Center}
  \city{Hannover}
  \country{Germany}}
\email{anand@l3s.de}

%
\renewcommand{\shortauthors}{}

%
\begin{abstract}

A recent trend in IR has been the usage of neural networks to learn retrieval models for text based adhoc search. While various approaches and architectures have yielded significantly better performance than traditional retrieval models such as BM25, it is still difficult to understand exactly \emph{why} a document is relevant to a query. In the ML community several approaches for explaining decisions made by deep neural networks have been proposed -- including DeepSHAP which modifies the DeepLift algorithm to estimate the relative importance (shapley values) of input features for a given decision by comparing the activations in the network for a given image against the activations caused by a reference input. In image classification, the reference input tends to be a plain black image. While DeepSHAP has been well studied for image classification tasks, it remains to be seen how we can adapt it to explain the output of Neural Retrieval Models (NRMs). In particular, what is a good ``black'' image in the context of IR? In this paper we explored various reference input document construction techniques. Additionally, we compared the explanations generated by DeepSHAP to LIME (a model agnostic approach) and found that the explanations differ considerably. Our study raises concerns regarding the robustness and accuracy of explanations produced for NRMs. With this paper we aim to shed light on interesting problems surrounding interpretability in NRMs and highlight areas of future work.



\end{abstract}

\copyrightyear{2019}
\acmYear{2019}
\setcopyright{acmcopyright}
\acmConference[SIGIR '19]{Proceedings of the 42nd International ACM SIGIR
Conference on Research and Development in Information Retrieval}{July 21--25,
2019}{Paris, France}
\acmBooktitle{Proceedings of the 42nd International ACM SIGIR Conference on
Research and Development in Information Retrieval (SIGIR '19), July 21--25, 2019,
Paris, France}
\acmPrice{15.00}
\acmDOI{10.1145/3331184.3331312}
\acmISBN{978-1-4503-6172-9/19/07}

%
%


%

%
\maketitle

\section{Introduction}

Deep neural networks have achieved state of the art results in several NLP and computer vision tasks in the last decade. Along with this spurt in performance has come a new wave of approaches trying to explain decisions made by these complex machine learning models. Explainablilty and interpretability are key to deploying NNs in the wild and having them work in tandem with humans. Explanations can help debug models, determine training data bias and understand decisions made in simpler terms in order to foster trust. Recently in IR, models such as DRMM~\cite{Guo16}, MatchPyramid~\cite{MatchPyramid}, PACRR-DRMM~\cite{pacrr_drmm} and others have shown great promise in ranking for adhoc text retrieval. While these models do improve state-of-the-art on certain benchmarks, it is sometimes hard to understand why exactly these models are performing better. With the increased scrutiny on automated decision making systems, including search engines, it is vital to be able to explain decisions made. In IR however, little to no work has been done on trying to explain the output of complex neural ranking models. 

In the ML community, several post-hoc non-intrusive methods have been suggested recently which enable us to train highly accurate and complex models while also being able to get a sense of their rationale. One of the more popular approaches to producing explanations is to determine the input feature attributions for a given instance and it's prediction according to a given model. The output of such a method is typically visualized as a heat map over the input words/pixels. Several approaches have been proposed in this direction for image and text classification but their applicability to adhoc text retrieval and ranking remains unexplored. In this paper we study the applicability of one such method designed specifically for neural networks -- DeepSHAP~\cite{Lundberg17}, to explain the output of 3 different neural retrieval models. DeepSHAP is a modification of the DeepLift~\cite{ShrikumarGK17} algorithm to efficiently estimate the shapley values over the input feature space for a given instance. The shapley value is a term coined by Shapley~\cite{shapley1953value} in cooperative game theory to refer to the contribution of a feature in a prediction. More specifically, shapley values explain the contribution of an input feature towards the \emph{difference} in the prediction made vs the average prediction value. 

The objective of our work is to utilize DeepSHAP to explain NRMs which should ideally be a trivial pursuit since they are standard neural networks. However, in our experiments, we found that DeepSHAP's explanations are highly dependent on a reference input which is used to compute the average prediction. This is inline with recent work that suggests approaches like DeepLIFT lack robustness~\cite{Ghorbani2017}. In this work, we ponder on the question, what makes a good reference input distribution for neural rankers? In computer vision, a plain black image is used as the reference input but what is the document equivalent of such an image in IR? Furthermore, we found that explanations produced by the model introspective DeepSHAP are considerably different from the model agnostic approach -- LIME~\cite{Ribeiro16}. Although both models produce local explanations, the variability is concerning. 



\section{Related Work}
\label{sec:rel_work}
There are two main approaches to interpretability in machine learning models: {\it model agnostic} and {\it model introspective} approaches. Model agnostic approaches \cite{Ribeiro16, Ribeiro18} generate post-hoc explanations for the original model by treating it as a black box by learning an interpretable model on the output of the model or by perturbing the inputs or both. Model introspective approaches on one hand include ``interpretable'' models such as decision trees \cite{letham2015}, attention-based networks \cite{Xu2015}, and sparse linear models \cite{Ustun2016} where there is a possibility to inspect individual model components (path in a decision tree, feature weights in linear models) to generate useful explanations. On the other hand, there are gradient-based methods like \cite{Simonyan2013DeepIC} that generates attributions by considering the partial derivative of the output with respect to the input features. Following this, there were many works \cite{Lundberg17, ShrikumarGK17, Bach15, Arras17} that generate attributions by inspecting the neural network architectures.  

{\bf Interpretability in ranking models} Recently there have been few works focused on interpretability \cite{Singh19, singh2018intentmodeling} and diagnosis of neural IR models \cite{Rennings19, PangLG0C17}. In the diagnostic approaches, they use the formal retrieval constraints (``axioms'') defined for traditional retrieval models to find the differences between neural IR and learning-to-rank approaches with hand-crafted features through a manual error analysis \cite{PangLG0C17} or build diagnostic datasets based on the axioms to empirically analyse these models \cite{Rennings19}. In \cite{Singh19} they built a explainable search system (EXS) that adapts a local model agnostic interpretability approach (LIME~\cite{Ribeiro16}) to explain the relevance of a document for a query for various neural IR models. 
In~\cite{singh2018intentmodeling} they propose an approach that understands the query intent encoded by NRMs by learning a simple ranking model with an expanded query that approximates the original ranking.

To the best of our knowledge, this is the first work that looks at model introspective interpretability specifically for NRMs.

\begin{figure*}[h]
  \centering
  \scalebox{0.82}{
\includegraphics[width=.33\linewidth]{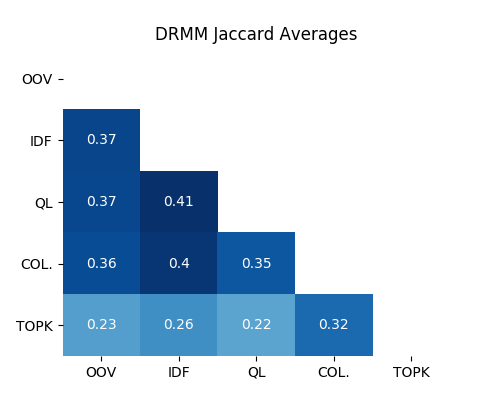}\hfill
\includegraphics[width=.33\linewidth]{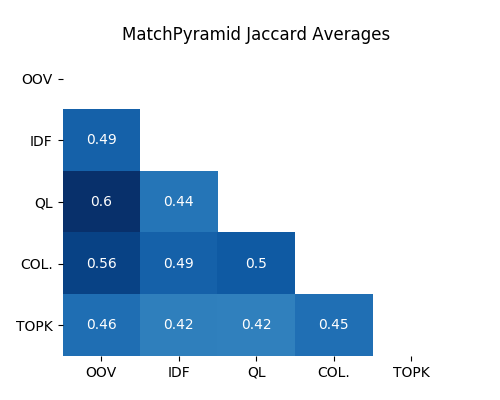}\hfill
\includegraphics[width=.33\linewidth]{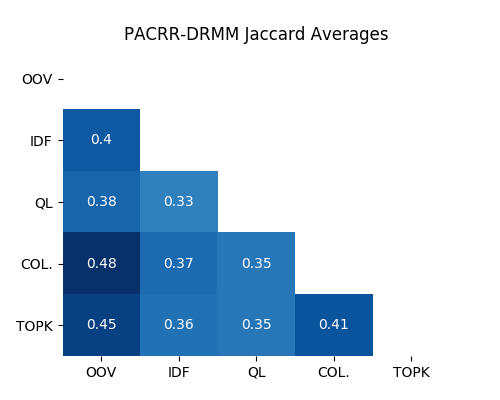}}
  \caption{Confusion matrices of various DeepSHAP background document methods comparing Jaccard similarities.}
  \Description{}
  \label{fig:lime_exp}
\end{figure*}

\section{DeepSHAP for IR}
\label{sec:shap}

DeepSHAP is a local model-introspective interpretability method to approximate the shapley values using DeepLIFT \cite{ShrikumarGK17}. DeepLIFT explains the difference in output/prediction from some `reference' output with respect to the difference of the input (to explain) from a `reference' input. The authors define a function analogous to partial derivatives to compute the feature importance scores and use the chain rule to backpropagate the activation differences from the output layer to the original input. The choice of reference input depends on domain specific knowledge; For example, in digit classification task on the MNIST dataset, they use a reference input of all-zeros as that is the background of the images. For object detection in images, a plain black image is often used.

In the context of IR, DeepSHAP can be used to explain why a document is relevant to query (according to a given NRM) by computing the shapley values for words in the document. 
The words with high shapley values indicate that they are important towards this prediction of relevance. However, to accurately compute the shapley values using DeepSHAP a reference input is needed. 
What makes a good background image in the context of IR? 

Unlike classification tasks, in ranking we have at least 2 inputs which are in most cases the query and document tokens. In this work we fix the reference input for the query to be same as that of the query-document instance to be explained and experiment with various reference inputs for the document. The intuition behind doing so is to gain an average reference output in the locality of the query. 

The various document reference inputs that we considered in our experiments are:
\begin{description}
    \item[OOV] The reference document consists of `OOV' tokens. For, DRMM and MatchPyramid models the embedding vector for `OOV' comprises of all-zeros which is similar to the background image used for MNIST. But for PACRR-DRMM,
    it is the average of all the embedding vectors in the vocabulary. 
    \item[IDF {\it lowest}] The reference document is constructed by sampling words with low IDF scores. These words are generally stop-words or words that are similar to stop-words so they should, in general, be irrelevant to the query. 
    \item[QL {\it lowest}] The reference document comprises of sampled words with low {\it query-likelihood} scores that are derived from a language model of the {\it top-1000} documents. 
    \item[COLLECTION {\it rand doc}] The reference document is randomly sampled from the rest of the collection minus the {\it top-1000} documents retrieved for the query.
    \item[TOPK LIST {\it rand doc from bottom}] The reference document is randomly sampled from the bottom of the {\it top-1000} documents retrieved.
\end{description}

These variants were designed based on the intuition that the reference input document would comprise of words that are irrelevant to the query and thus DeepSHAP should be able to pick the most important terms from the input document that explain relevance to the query.

\section{Experimental Setup}
\label{sec:experiments}

In our experiments, we aim to answer the following research questions:

\begin{itemize}
    \item Are DeepSHAP explanations sensitive to the type of reference input in the case of NRMs?
    \item Can we determine which reference input produces the most accurate local explanation?
\end{itemize}

To this end, we describe the experimental setup we used to address these questions. We describe the various NRM's we considered and how we used LIME to evaluate the correctness of explanations produced by DeepSHAP.

\subsection{Neural Retrieval Models}

    {\bf DRMM} \cite{Guo16} This model uses a query-document term matching count histogram as input into a feed forward neural network (MLP) to output a relevance score along with a gating mechanism that learns query term weights.
    
    {\bf MatchPyramid} \cite{MatchPyramid} This model uses a query-document interaction matrix as input to a 2D CNN to extract matching patterns. The output is then fed into a MLP to get a relevance score.
    
    {\bf PACRR-DRMM} \cite{pacrr_drmm} This model creates a query-document interaction matrix that is fed into multiple 2D CNNs with different kernel sizes to extract {\it n}-gram matches. Then, after {\it k}-max pooling across each q-term, 
    the document aware q-term encodings are fed into a MLP, like in DRMM to obtain a relevance score.

\subsection{Evaluation}

To conduct the experiments, we used the Robust04 test collection from TREC. We used Lucene to index and retrieve documents. Next, we trained the NRMs using their implementations in MatchZoo\footnote{\url{https://github.com/NTMC-Community/MatchZoo/tree/1.0}}~\cite{FanPHGLC17}. All the hyparameters were tuned using the same experimental setup as described in the respective papers. We chose to study explanations for the distinguished set of hard topics\footnote{\url{https://trec.nist.gov/data/robust/04.guidelines.html}} (50) from the TREC Robust Track 2004. We generate the explanations from LIME and DeepSHAP for the top-3 documents retrieved for each query and use only these for our quantitative experiments. 

\textbf{Evaluating explanations} Since no ground truth explanations are available for a neural model, we use LIME based explanations as a proxy. 
We found that it can approximate the locality of a query document pair well, using a simple linear model that achieved an accuracy of over 90\% across all NRMs considered in our experiments.
To produce the explanations from LIME we used the implementation found in \footnote{\url{https://github.com/marcotcr/lime}} along with the score-based modification suggested in~\cite{Singh19}. The primary parameters for training a LIME explanation model are the number of perturbed samples to be considered and the number of words for the explanation. The number of perturbed samples is set to 5000 and the number of words is varied based on the experiment. We used the DeepSHAP implementation provided here~\footnote{\url{https://github.com/slundberg/shap}}. Note that we ignore the polarity of the explanation terms provided by both methods in our comparison since the semantics behind the polarities in LIME and DeepSHAP are different. We are more interested in the terms chosen as the explanations in both cases.


\begin{table*}
\footnotesize
\caption{Comparison of recall measures at \textit{top-k} (50, 100) terms from DeepSHAP without polarity against the \textit{top-k} (10, 20, 30) ground-truth terms from LIME for ROBUST04 difficult queries (50)}
\label{tab:doc_bg_dist_recall}
\begin{tabular}{ m{6em}m{2em}m{2em}m{2em}m{2em}m{2em}m{2em}m{2em}m{2em}m{2em}m{2em}m{2em}m{2em}m{2em}m{2em}m{2em}m{2em}m{2em}m{2em} } 
\toprule
 &\multicolumn{6}{c}{DRMM}&\multicolumn{6}{c}{MatchPyramid}&\multicolumn{6}{c}{PACRR-DRMM}\\
 \cmidrule(r){2-7}
 \cmidrule(r){8-13}
 \cmidrule(r){14-19}
 &\multicolumn{2}{c}{top-10} &\multicolumn{2}{c}{top-20}&\multicolumn{2}{c}{top-30} &\multicolumn{2}{c}{top-10}&\multicolumn{2}{c}{top-20}&\multicolumn{2}{c}{top-30}
 &\multicolumn{2}{c}{top-10}&\multicolumn{2}{c}{top-20}&\multicolumn{2}{c}{top-30}\\
 \cmidrule(r){2-3}
 \cmidrule(r){4-5}
 \cmidrule(r){6-7}
 \cmidrule(r){8-9}
 \cmidrule(r){10-11}
 \cmidrule(r){12-13}
 \cmidrule(r){14-15}
 \cmidrule(r){16-17}
 \cmidrule(r){18-19}
 SHAP \newline variants & recall\newline @50  & recall\newline @100 & recall\newline @50  & recall\newline @100 & recall\newline @50  & recall\newline @100 & recall\newline @50  & recall\newline @100 & recall\newline @50  & recall\newline @100 & recall\newline @50  & recall\newline @100 & recall\newline @50  & recall\newline @100  & recall\newline @50  & recall\newline @100 & recall\newline @50  & recall\newline @100\\
 \midrule

OOV & 0.789 & 0.905 & 0.672 & 0.845 & 0.615 & 0.812 & {\bf 0.793} & {\bf 0.843} & {\bf 0.656} & {\bf 0.726} & {\bf 0.566} & {\bf 0.640} & 0.582 & 0.582 & 0.388 & 0.388 & 0.299 & 0.299\\
\cmidrule(lr){2-19}

IDF & 0.830 & 0.917 & 0.723 & 0.871 & 0.658 & 0.841 & 0.795 & 0.832 & 0.653 & 0.711 & 0.565 & 0.633 & 0.633 & 0.633 & 0.446 & 0.446 & 0.362 & 0.362\\
\cmidrule(lr){2-19}

QL & {\bf 0.894} & {\bf 0.955} & {\bf 0.754} & {\bf 0.892} & {\bf 0.670} & {\bf 0.856} & 0.765 & 0.821 & 0.638 & 0.711 & 0.556 & 0.636 & {\bf 0.643} & {\bf 0.643} & {\bf 0.462} & {\bf 0.462} & {\bf 0.367} & {\bf 0.367}\\
\cmidrule(lr){2-19}

COLLECTION & 0.760 & 0.881 & 0.673 & 0.841 & 0.620 & 0.815 & 0.783 & 0.824 & 0.639 & 0.709 & 0.552 & 0.630 & 0.621 & 0.621 & 0.429 & 0.429 & 0.343 & 0.343\\
\cmidrule(lr){2-19}

TOPK LIST. & 0.639 & 0.821 & 0.606 & 0.794 & 0.578 & 0.788 & 0.759 & 0.811 & 0.624 & 0.702 & 0.545 & 0.627 & 0.625 & 0.625 & 0.425 & 0.425 & 0.340 & 0.340\\
\bottomrule

 \end{tabular}
\end{table*}

\begin{table}
\footnotesize
\caption{Comparison of mean squared error (MSE) and accuracy (ACC) of LIME's linear model across various NRMs. Low MSE and high accuracy shows that it is able to fit and generalize in the query-document locality.}
\label{tab:lime_model_performances}
\scalebox{0.85}{
\begin{tabular}{lcccc} 
 \toprule
 &\multicolumn{4}{c}{Linear Regression}\\
 NRM & TRAIN MSE & TEST MSE & TRAIN ACC & TEST ACC\\
 \midrule
 DRMM & 0.00631 & 0.00633 & 0.92662 & 0.92654\\
 MatchPyramid & 0.01827 & 0.01839 & 0.90367 & 0.90387\\
 PACRR-DRMM & 0.00165 & 0.00160 & 0.98857 & 0.98980\\
\bottomrule
\end{tabular}}
\end{table}

\begin{table}
\caption{An example of words selected by LIME and SHAP methods for the query {\it `cult lifestyles'} and document {\it `FBIS3-843'} which is about clashes between cult members and student union's activists at a university in Nigeria. {\it Words unique to a particular explanation method are highlighted in bold}.}
\centering
\scalebox{0.77}{
\begin{tabular}{cccccc}
\toprule
LIME & OOV & IDF & QL & COL. & TOPK\\
\midrule
cult & cult & cult & cult & cult & cult \\
style & followers & style & style & black & {\bf numbers} \\
followers & black & followers & elite & fraternities & {\bf english} \\
elite & fraternities & {\bf suspects} & saloon & degenerate & {\bf college} \\
saloon & degenerate & {\bf belong} & {\bf final} & sons & {\bf university} \\
{\bf student} & sons & {\bf reappearing} & march & followers & {\bf fallouts} \\
home & {\bf academic} & household & {\bf friday} & style & {\bf buccaneers} \\
{\bf members} & {\bf american} & black & september & home & {\bf feudings} \\
march & {\bf tried} & fraternities & {\bf arms} & household & {\bf activists} \\
september & household & degenerate & {\bf closed} & {\bf avoid} & {\bf troubles} \\

\bottomrule
\end{tabular}
}
\label{tab:qualitative_examples}
\end{table}

\section{Results and Discussion}

\subsection{Effect of reference input document}

Figure~\ref{fig:lime_exp} illustrates the overlap between the explanation terms produced when varying the reference input. Immediately we observe that the overlap between explanations produced is low; below 50\% in most cases and consistently across all NRMs. Each reference input method has its own distinct semantics and this is reflected by the low overlap scores. We also find that there is no consistent trend across NRMs. 
For MatchPyramid, OOV and QL have highest overlap whereas for PACRR-DRMM its OOV and COL that have highest overlap even though both models share similarities in input representation and parts of the model architecture. Table~\ref{tab:qualitative_examples} shows explanations for DRMM across variants. Once again we see how the explanations can differ significantly if we are not careful in selecting the reference input. For IR, finding the background image seems to be a much harder question. 

Our results show how explanations are highly sensitive to the reference input for NRMs chosen in our experiments. This is also indication that a single reference input method may not be the best for every NRM. 

\subsection{Accuracy of reference input methods}

To help identify which reference input method is most accurate in explaining a given query-document pair, we compared the LIME explanations for the same against it's corresponding DeepSHAP explanations. In general we found that DeepSHAP produces more explanation terms whereas LIME's L1 regularizer constrains the explanations to only the most important terms. Additionally, the discrepancy between the explanations can be attributed to LIME being purely local, whereas DeepSHAP has some global context since it looks at activations for the whole network which may have captured some global patterns. Hence, to estimate which reference input surfaces the most `ground truth' explanation terms we only computed recall at top 50 and 100 (by shapley value magnitude) DeepSHAP explanation terms (in Table~\ref{tab:doc_bg_dist_recall}).

The first interesting insight is that some NRMs are easier to explain whereas others are more difficult. PACRR-DRMM consistently has a recall less than $0.7$, whereas the DeepSHAP explanations of DRMM effectively capture almost all of the LIME explanation terms. When comparing reference input variants within each NRM we find that there is no consistent winner. For DRMM, QL is the best which indicates that sampling terms which are relatively generic for this query in particular is a better `background image' than sampling generic words from the collection (IDF). 

In the case of MatchPyramid, TOPK LIST is the worst performing but it is more difficult to distinguish between the approaches here. The best approach surprisingly is OOV. This can be attributed to how MatchPyramid treats OOV terms. The OOV token is represented by an all-zeros embedding vector that is used for padding the input interaction matrix whereas in DRMM, OOV tokens are filtered out. These preprocessing considerations prove to be crucial when determining the right input reference. Moving on to PACRR-DRMM, we once again find that QL is the best method even though DeepSHAP struggles to find all the LIME terms.

\section{Conclusion and Future Work}

In this paper we suggested several reference input methods for DeepSHAP that take into account the unique semantics of document ranking and relevance in IR. Through quantitative experiments we found that it is indeed sensitive to the reference input. The distinct lack of overlap in most cases was surprising but in line with recent works on the lack of robustness in interpretability approaches. We also tried to evaluate which reference method is more accurate by comparing against LIME. Here we found that reference input method selection is highly dependent on the model at hand. We believe that this work exposes new problems when dealing with model introspective interpretability for NRMs. A worthwhile endeavor will be to investigate new approaches that explicitly take into account the discreteness of text and the model's preprocessing choices when generating explanations.

This work was supported by the Amazon research award on `Interpretability of Neural Rankers'.




%

%
\bibliographystyle{ACM-Reference-Format}
\bibliography{sample-base}

%









\end{document}